\documentclass[times, twoside, watermark]{zHenriquesLab-StyleBioRxiv}
\usepackage{subcaption}

\leadauthor{Platt} 

\begin{document}

% \subtitle{Bioimage informatics}

\title{
    MiCellAnnGELo: Annotate microscopy time series of complex cell surfaces with 3D Virtual Reality}
\shorttitle{MiCellAnnGELo}
\author[1]{
    Adam Platt
}
\author[1,\Letter]{    
    E. Josiah Lutton 
}
\author[1]{
    and Till Bretschneider
}
\affil[1]{
    Department of Computer Science,
    University of Warwick, 
    Coventry, 
    CV4 7AL, 
    United Kingdom
}

\maketitle

\abstract{
    \textbf{Summary:} 
    Advances in 3D live cell microscopy are enabling high-resolution capture of previously unobserved processes. 
    Unleashing the power of modern machine learning methods to
    fully benefit from these technologies
    is, however, frustrated by the difficulty of manually annotating 3D training data. 
    MiCellAnnGELo virtual reality software offers an immersive environment for viewing and interacting with 4D microscopy data, including efficient tools for annotation. 
    We present tools for labelling  cell surfaces with a wide range of applications, including cell motility, endocytosis,  and transmembrane signalling.\\
    \textbf{Availability and implementation:} 
    MiCellAnnGELo employs the cross platform (Mac/Unix/Windows) Unity game engine and is available under the MIT licence at https://github.com/CellDynamics/MiCellAnnGELo.git, together with sample data and demonstration movies. 
    MiCellAnnGELo can be run in desktop mode on a 2D screen or in 3D using a standard VR headset with compatible GPU.
}

\begin{corrauthor}
\href{josiah.lutton@warwick.ac.uk}{josiah.lutton@warwick.ac.uk}
\end{corrauthor}

\section{Introduction}
In recent years, artificial intelligence (AI) has become a powerful tool for bioimage analysis. 
A major roadblock for the application of AI is obtaining manually annotated training data, which requires human experts to label image features on a computer screen, a tedious process prone to a high error rate. 
Standard approaches for labelling 3D objects use 2D cross-sections to view and annotate the volumes,
which, because single 3D objects can appear disjointed in 2D sections,
requires excessive input from the annotator, limiting the number of annotations produced.
We aim to address this issue with our software, MiCellAnnGELo (\textbf{Mi}croscopy and \textbf{Cell} \textbf{Ann}otation \textbf{G}raphical \textbf{E}xperience and \textbf{L}abelling to\textbf{o}l), an easy-to-use virtual reality (VR) interface for annotating dynamic biological surfaces.

The use of VR for data visualization dates back over 25 years \citep[for example][]{fruhauf1996}. 
Recent advances in graphics processing unit (GPU) and VR headset technology have greatly expanded the range of applications of this technology. 
ConfocalVR~\citep{stefani2018}, 
Arivis VisionVR~\citep{conrad2020}, and syGlass~\citep{pidhorskyi2018} are examples of VR software for annotating 3D biological image volumes.
SlicerVR~\citep{pinter2020} is a VR visualization plugin for the open-source software 3D~Slicer~\citep{fedorov2012}.
TeraVR~\citep{wang2019} provides a specialized VR application for neuron tracing in 3D image volumes, with functionality for placing markers and surface visualization, as part of the open-source software Vaa3D~\citep{peng2010}.
ChimeraX~\citep{pettersen2021} is primarily focused on molecular visualization and analysis, but has also been applied to biological image analysis and surface visualization~\citep{driscoll2019,quinn2021}.

MiCellAnnGELo aims to facilitate fast annotation of 3D microscopy movies.
We have focused development on annotation of time series of cell surfaces, 
which is a less computationally intensive task than directly interacting with the 3D microscopy movies,
allowing annotation of large movies without requiring high GPU specifications.
This is made possible by recent advances in cell segmentation methods
\citep[e.g.][]{arbelle2022,eschweiler2022,lutton2021}
allowing the extraction of the cell surface meshes.
The example meshes used in the following additionally contained fluorescence data, taken from the source image
using local maximum fluorescence \citep[see][for details]{lutton2022}.
Two annotation methods are available in MiCellAnnGELo:
a ``mesh painting'' feature that allows fast labelling of the surfaces for a whole time series, 
while markers can be placed for feature tracking.
We are releasing the software as an open-source tool to facilitate use and development within the community.
Finally, MiCellAnnGELo is designed for VR visualization, allowing development to focus on streamlining
this environment for annotation.

\section{Software Features}
\label{sec:software}
MiCellAnnGELo is a cross-platform software program that provides an immersive environment for the rapid annotation of series of triangulated surface meshes.
Meshes with single- or dual-channel colour mappings can be displayed and annotated in the environment.
The software provides both VR and desktop interfaces, with easy interchange between these modes.

\begin{figure*}
    \centering
    \includegraphics[width=\linewidth]{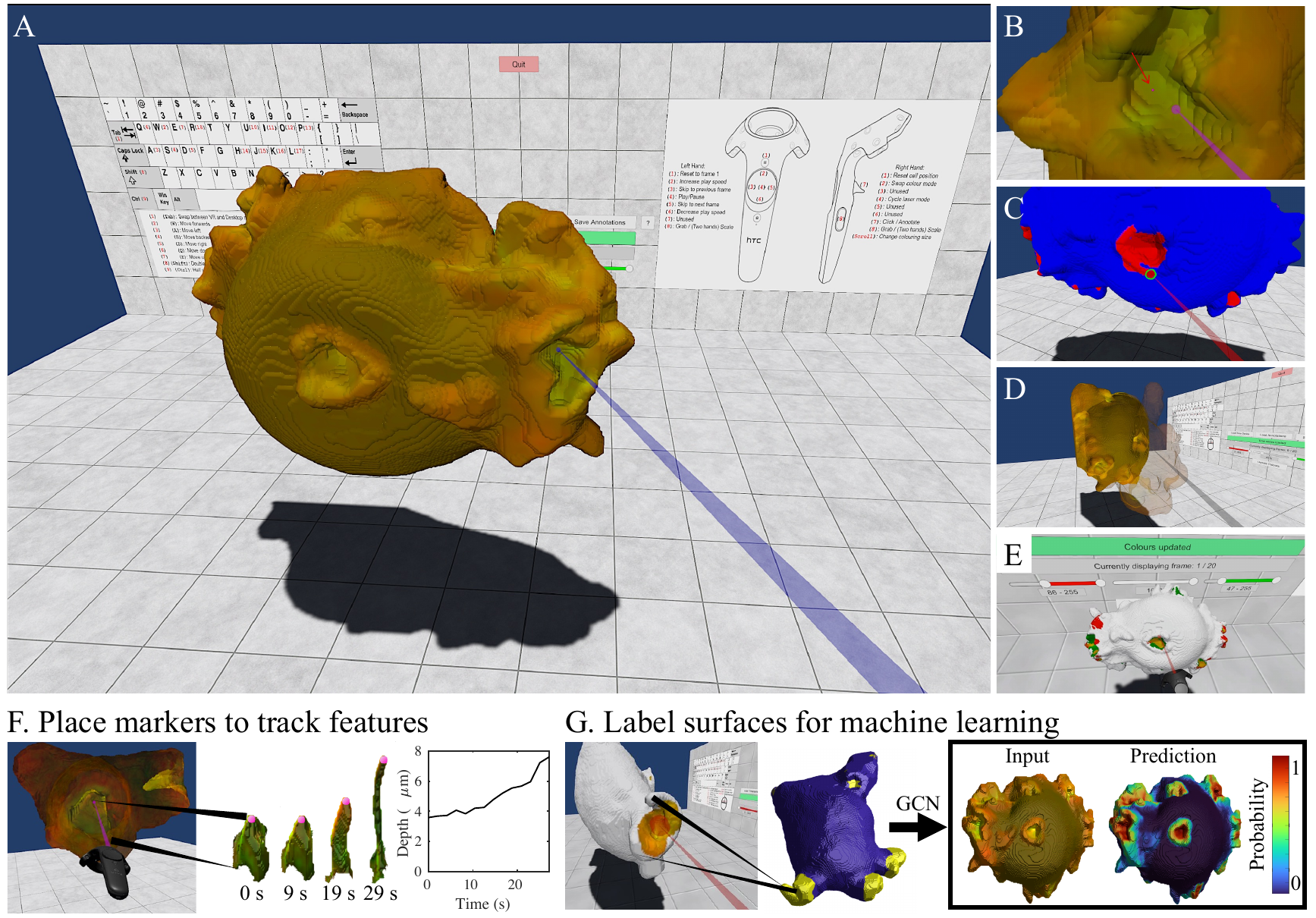}
    \caption{
    \textbf{A}:~Overview of the MiCellAnnGELo environment. 
    \textbf{B}:~Placement of a marker (red arrow) using the laser pointer. 
    \textbf{C}:~Mesh painting in the two-tone annotation mode using the size-adjustable paint tool (green outline).
    \textbf{D}:~Transparency of part of the surface can be adjusted using the laser pointer.
    \textbf{E}:~Identifying colour-dependent features can be facilitated through colour thresholding.
    \textbf{F}:~One application of the software is the ability to rapidly place markers (left) in a sequence of surfaces, allowing manual tracking of surface features (middle), which can be paired with a feature detection method to allow time-dependent measurements to be made (right).
    \textbf{G}:~A second application of the software is to generate training data by painting labels on the surface (left), which can be used by a machine learning algorithm, e.g. a graph convolutional neural network (GCN) to predict features from input labels (right panel).
    }
    \label{fig1}
\end{figure*}
% \subsection{User interface}
%\label{sec:ui}
% \textit{User interface} 
The user interface is designed with simplicity in mind, allowing the user to load, explore, and annotate data with ease. 
As can be seen in Fig~\ref{fig1}\textbf{A}, the environment consists of the surface mesh itself and a wall-mounted user interface, which provides controls for loading and saving data, and for adjusting surface colour and opacity.
Additionally, controller layouts for both VR and desktop modes are displayed on the wall for ease of use.
In both VR and desktop environments, functions including changing annotation modes or surface representations, moving forwards/backwards in time, and playing/pausing the series are mapped to single buttons, enabling rapid multi-modal annotation of a sequence.

% \subsection{Annotation}
% \label{sec:annotation}
% \textit{Annotation}
Annotations can be made by placing markers (Fig~\ref{fig1}\textbf{B}, Supplementary Video~1) and by mesh painting (Fig~\ref{fig1}\textbf{C}, Supplementary Video~2).
Marker placement allows features
to be marked with a single shot per frame, 
allowing rapid manual tracking of surface features across multiple frames.
Painting allows rapid labelling of larger structures on the surface, with an easily adjustable brush size and eraser mode.
The painted surface can be visualized either as a two-tone image or as a cut-out image with  mesh colours being blocked out in unlabelled areas  (Fig~\ref{fig1}\textbf{G} left, Supplementary~Video~2).

% \subsection{View controls}
% \label{sec:view}
% \textit{View controls}
A number of view controls have been added to enable easier labelling of data.
In VR mode, the user can freely adjust the surface mesh position and size (Supplementary Video~3),
allowing the user to rapidly change perspectives.
For annotating more complex surfaces, the user can
reduce the opacity of parts of the surface 
(Fig~\ref{fig1}\textbf{D}, Supplementary Video~4).
Finally, the mesh colours can be adjusted using controls on the wall UI (Fig~\ref{fig1}\textbf{E}, Supplementary Video~4),
allowing increased visibility of colour-dependent features.

% \subsection{Input and output files}
% \label{sec:io}
% \textit{Input and output files}
Surface meshes can be loaded into the environment from a sequence of .ply files.
This format stores colour data, and can be generated from surfaces in 
% R1 wants examples of how these can be generated | Change made
many image analysis software applications~(e.g. python via PyVista~\citep{sullivan2019} and Matlab via plywrite~\citep{cecen2015}).
MiCellAnnGELo takes the first two channels of the .ply files as colour data.
Marker annotations are exported as .csv files encoding
the frame number, spatial coordinates, and vertex index
in the surface mesh of each marker 
(example shown in Supplementary Video 1).
Painted surfaces are exported as .ply files,
keeping the original colours in the first two channels and
placing the paint labels in the third channel, and can
be retrieved in future sessions by selecting this sequence in the load menu.
These labels can be read (e.g. via PyVista~\citep{sullivan2019}, Meshlab~\citep{cignoni2008}, 3D~Slicer~\citep{fedorov2012}, and Matlab via plyread~\citep{getreur2004}) for further analysis in other software by extracting the blue channel.

We demonstrate two example use cases in Fig~\ref{fig1}\textbf{F--G}.
In both cases, annotations were made on the surface of a cell undergoing macropinocytosis (cell drinking),
with the aim to identify the concave structures, referred as cups, that are associated with this process.
In the first, we used the marker placement tool to mark the centres of individual cups over time.
Applying a threshold to the green channel gives an approximate shape of the marked cup in each frame,
allowing time-dependent geometric variation of the cups to be measured.
Using the mesh painting tool to label the cups enables
machine learning methods to be applied, yielding a more accurate representation of the cups.
The labels were used as training data for a graph convolutional neural network,
which could then isolate the cups in a much larger set of surfaces.

\section{Conclusion}
MiCellAnnGELo is a cross-platform open-source software application designed to allow rapid annotation of surface series using VR technology.
The software is streamlined and easy to use, with a range of viewing and annotation options.
In future developments we aim to introduce functionality for 3D volumetric data and multi-object series, and allow more input and output formats.
We aim to develop this software as a community resource project, allowing further development to be geared towards providing a VR software solution for a wide range of microscopy use cases and easy integration into existing pipelines.

\section*{Acknowledgements}
\textbf{Funding:} 
This work was supported by Biotechnology and Biological Sciences Research Council [BB/R004579/1 to T.B.] and Engineering and Physical Sciences Research Council [EP/V062522/1 to E.J.L, T.B.].

\bibliography{vr}

\end{document}